\documentclass[12pt]{article}

\usepackage[margin=3cm]{geometry}
\usepackage{graphicx}
\usepackage{hyperref}

\usepackage{amsmath, amssymb}
\usepackage{amsthm}
\usepackage{mathrsfs} % voor Lagrangiaan

\newcommand{\calL}{\mathcal{L}}

\newcommand{\bfG}{\mathbb{G}}

\newcommand{\bfR}{\mathbb{R}}

\newcommand{\bfZ}{\mathbb{Z}}

\newcommand{\tr}{\mathrm{tr}}

\newcommand{\dual}[1]{\left<{#1}\right>}

\newcommand{\Ad}{\mathrm{Ad}}

\newcommand{\dd}[1]{\frac{\d}{\d #1}}
\newcommand{\ddt}{\dd{t}}
\newcommand{\dde}{\dd{\epsilon}}

\newcommand{\lcf}{\lbrack\!\lbrack} % math.DG/0506299
\newcommand{\rcf}{\rbrack\!\rbrack}

\newcommand{\AS}[1]{\lcf{#1}\rcf}

\newcommand{\contraction}{\vrule height 0pt depth 0.4pt width 3pt
  \vrule height 7pt depth 0.4pt \kern 3pt}
\newcommand{\be}{\begin{equation}}
\newcommand{\ee}{\end{equation}}

\ifx\isjournal\undef
\newtheoremstyle{jvk-thm} %
  {}{}{\itshape}{}{\bfseries}{.}{ }{}
\newtheoremstyle{jvk-rem} %
  {}{}{\upshape}{}{\bfseries}{.}{ }{}

\theoremstyle{jvk-thm}
\newtheorem{definition}{Definition}[section]
\newtheorem{lemma}[definition]{Lemma}
\newtheorem{theorem}[definition]{Theorem}
\newtheorem{prop}[definition]{Proposition}

\newenvironment{mproof}{\textbf{Proof:}\,}{\hfill$\diamond$}

\theoremstyle{jvk-rem}
\newtheorem{exampleth}[definition]{Example}
\newenvironment{example}{\begin{exampleth}}{\hfill$\diamond$\end{exampleth}} 

\newcommand{\myremarksymbol}{$\diamond$}
\newtheorem{remarkth}[definition]{Remark}
\newenvironment{remark}{%
   \begin{remarkth}\pushremarksymbol
}{\popremarksymbol\end{remarkth}} 

\newcommand{\pushremarksymbol}{%
  \makeatletter
    \global\def\@remcommand{\hfill\myremarksymbol}
  \makeatother
}
\newcommand{\popremarksymbol}{%
  \makeatletter
    \@remcommand\global\def\@remcommand{\relax}
  \makeatother
}
\fi

\newcommand{\define}[1]{\textbf{#1}}
\newenvironment{AlphaList}{%
  \begin{enumerate}}{\end{enumerate}}

\newcommand{\arxiv}[1]{\texttt{#1}}
\newcommand{\gothe}{\mathfrak{e}}
\newcommand{\gothf}{\mathfrak{f}}
\newcommand{\gothg}{\mathfrak{g}}

\newcommand{\gl}{\mathfrak{gl}}

\renewcommand{\d}{\mathrm{d}}

\newcommand{\qforall}{\quad \text{for all}\:}

\newcommand{\qqand}{\quad \mathrm{and} \quad}

\title{Euler-Poincar\'e reduction for discrete field theories}
\date{}

\begin{document}
\maketitle

\bigskip

Joris Vankerschaver\\
\textit{{\small% 
Department of Mathematical Physics and Astronomy \\
University of Ghent, Krijgslaan 281, \\
B-9000 Ghent, Belgium}} \\
e-mail: \url{Joris.Vankerschaver@UGent.be}
\bigskip

\begin{abstract}
  In this note, we develop a theory of Euler-Poincar\'e reduction for
  discrete Lagrangian field theories.  We introduce the concept of
  Euler-Poincar\'e equations for discrete field theories, as well as a
  natural extension of the Moser-Veselov scheme, and show that both
  are equivalent.  The resulting discrete field equations are
  interpreted in terms of discrete differential geometry.  An
  application to the theory of discrete harmonic mappings is also
  briefly discussed.
\end{abstract}

\bigskip
MSC classification (2000): 58H05; 65P10; 70S10\\

\clearpage

\section{Introduction}

Over the past decades, the interest in discrete mechanics has been
steadily increasing.  Partly this is due to the development of
geometric integrators, which provide a relatively simple means for the
long-term numerical integration of mechanical systems.  On the other
hand, many concepts (Poisson tensors, reduction, etc.) from the
continuum theory also have a natural counterpart in the discrete
realm, which makes discrete mechanics a subject worthy of interest in
its own right.  We refer here in particular to the pioneering efforts
of Moser and Veselov \cite{MoserVeselov91}, who studied a class of
discrete integrable systems, and Weinstein \cite{Weinstein96}, who was
the first to recognize the fact that many of these discrete systems
can be understood from the point of view of Lie groupoids.

Inspired by these powerful and elegant methods, a number of people set
out to develop a similar geometric approach to classical field
theories.  Bridges \cite{Bridges97} introduced a concept of
``multisymplecticity'' for Hamiltonian partial differential equations
and later Bridges \& Reich \cite{Bridges01} studied numerical
integrators that conserve a discretized version of multisymplecticity.
Independently, Marsden, Patrick, and Shkoller \cite{MPS98} extended
the work of Veselov in order to deal with Lagrangian field theories.
In a previous paper \cite{discreet}, we built upon their work by
extending the Lie-groupoid methods from Weinstein's paper and related
work by Marrero, Mart\'{\i}n, and Mart\'{\i}nez \cite{GroupoidMech05}
to the case of Lagrangian field theories, allowing us to treat a
larger class of field theories, as well as shedding new light on some
of the constructions in \cite{MPS98}.

In this note, we now focus on a special class of discrete Lagrangian
field theories and their behaviour in the presence of symmetry: we
introduce a reduction procedure, similar to the one proposed by
Marsden, Pekarsky, and Shkoller \cite{MarsdenPekarskyShkoller99} and
Bobenko and Suris \cite{BobenkoSuris99CMP, BobenkoSuris99LMP} for
discrete mechanical systems, and show that the resulting discrete
field equations have a simple and natural interpretation in the
context of discrete geometry.

We begin by introducing a fixed mesh (a certain collection of
vertices, edges, and faces) in the space of independent variables.
Initially, we are interested in discrete fields that associate to each
vertex an element of a given Lie group $G$.  If the Lagrangian of such
a model is $G$-invariant, we show in \S\ref{sec:DEP} that these fields
can be reduced to a new class of fields that associate a group element
to each \emph{edge} of the given mesh.  This reduction procedure is a
field-theoretic version of the discrete Euler-Poincar\'e equations
from \cite{MarsdenPekarskyShkoller99}.  We then show that the converse
procedure (of reconstruction) is possible only if a certain
obstruction vanishes.  This obstruction has a natural interpretation
as the curvature of a discrete $G$-connection.  This is similar to the
case of continuum field theories, as studied in \cite{MarcoLMP,
  MarcoRatiu}.

In \S\ref{sec:noether}, we take an alternative route to the
Euler-Poincar\'e equations: inspired by a similar treatment in
\cite{MarcoLMP}, we prove a discrete version of Noether's theorem and
use the $G$-invariance of the Lagrangian to derive the field
equations.  Finally, in \S\ref{sec:MV}, we consider the Lagrangian of
harmonic mappings into a Lie group: we propose an extension of the
well-known Moser-Veselov algorithm, demonstrate its equivalence to the
Euler-Poincar\'e equations, and establish a particularly clear form of
the field equations (involving the concepts of discrete geometry
introduced in \S\ref{sec:discDG}).

Throughout the paper, a number of simplifying assumptions will be
made.  In \S\ref{sec:general} we therefore conclude this paper by
giving a brief overview of how these assumptions may be circumvented,
and we also indicate how the discrete field theories in this paper fit
into the framework of Lie groupoid field theories developed in
\cite{discreet}.  In particular, we show that the Euler-Poincar\'e
reduction procedure in \S\ref{sec:DEP} is just a particular instance
of a general reduction theorem in the category of Lie groupoid field
theories.

\section{Discrete differential geometry} \label{sec:discDG}

\subsection{Discretizing the base space} \label{sec:mesh}

% \subsection{Discretizing the base space}

In this paper, we consider field theories with two independent
variables.  Such field theories can be modeled as sections of a fibre
bundle with base space $\bfR^2$ (see \cite{MPS98}).  In order to
discretize these fields, we introduce the \emph{set of vertices} $V$
in $\bfR^2$ as the set of points in $\bfR^2$ with integer coordinates
($V = \bfZ \times \bfZ$).  As in finite-difference approximations, the
idea is that continuous fields are approximated by specifying their
values on the elements of $V$.

For future reference, we also introduce the \emph{set of edges} $E$,
whose elements are ordered pairs of the following form: for all $i, j
\in \bfZ \times \bfZ$, 
\[
  ((i,j), (i+1, j)) \in E, \qqand ((i, j), (i, j+1)) \in E.
\]
Furthermore, we also demand that if $(x_0, x_1) \in E$, then also
$(x_1, x_0) \in E$, and we write $(x_1, x_0) = (x_0, x_1)^{-1}$.  In
other words, $E$ consists of ``horizontal'' and ``vertical'' line
segments of unit length.  To avoid cumbersome notation, we will note
an arbitrary element of $E$ as $\gothe$, or if we have to refer to its
begin and end vertex, as $(x_0, x_1)$.

Finally, we introduce the \emph{set of faces} $F$ as the set of
quadruples $(\gothe_0, \gothe_1, \gothe_2, \gothe_3)$ of closed paths
in $E$, i.e. where the end vertex of $\gothe_i$ is the begin vertex of
$\gothe_{i + 1}$ for $i = 0, \ldots, 3$, and, in addition, we demand
that $\gothe_i \ne \gothe_{i + 1}^{-1}$ for $i = 0, \ldots, 3$.  (The
indices should be interpreted as being ``modulo $4$'': $\gothe_4$ is
just $\gothe_0$.)

The elements of $F$ can thus be pictured as little squares in
$\bfR^2$.  A generic element of $F$ will be noted as $\gothf$, or,
since $\gothf$ is fully determined by its bounding vertices, as $(x_0,
x_1, x_2, x_3)$.  Note that there are two distinct classes of elements
in $F$: those where the vertices are denoted in clockwise and
anticlockwise fashion, respectively.

\subsection{Discrete differential forms} \label{sec:discform}

This section is devoted to a review of some elementary concepts of
algebraic topology, which will be needed in our discussion of discrete
harmonic maps in section~\ref{sec:MV}.  Most of this section is
modeled on \cite{DEC}, as well as on the text book \cite{Hatcher}.  

Consider the mesh $(V, E)$ in $\bfR^2$ introduced in the previous
section.  The collection of sets $\{V, E, F\}$ together with its
various incidence relations determines a cell complex and this
leads us naturally to the concepts of homology and cohomology.  It is
therefore not unreasonable to expect that some of these concepts will
enter our study of discrete field theories later on.

An \emph{$n$-chain} on the given mesh, where $n = 0, 1, 2$, is a
formal linear combination (with coefficients in $\bfR$) of
``$n$-dimensional elements''.  More precisely, the vector space of
$0$-chains is given by
\begin{equation} \label{defC0}
  C_0 = \{ \alpha_1 x_1 + \cdots + \alpha_m x_m : \alpha_1, \ldots,
  \alpha_m \in \bfR, x_1, \ldots, x_m \in V\},
\end{equation}
where it should be stressed that the elements of $C_0$ are
\emph{formal} linear combinations of elements of $V$.  Similarly, the
vector space $C_1$ of $1$-chains is generated by elements of $E$,
$C_2$ by elements of $F$, and so on.  Furthermore, for the space of
$1$-chains, we adopt the identification that $-(x_0, x_1) = (x_1,
x_0)$.

% \begin{remark} \label{remark:id} We recall that each edge in $E$ can
%   be represented by an oriented segment of a straight line in
%   $\bfR^2$, and hence is determined by its begin and end vertex.  This
%   is reflected in our notation by writing an edge $\gothe$ simply as
%   an ordered pair $(x_0, x_1)$.  
% \end{remark}

It is customary to define \emph{discrete $n$-forms} as $n$-dimensional
cochains, i.e. elements of the dual vector space $C_n^\ast$ of $C_n$.
From this definition, it follows immediately that a discrete zero-form
induces a function $\phi: V \rightarrow \bfR$.  Conversely, such a
function gives rise to a zero-form through linear extension.

Similarly, discrete one-forms can be identified with functions
$\varphi$ defined on the set of edges.  It should be borne in mind
that these functions are not necessarily defined on the whole of $V
\times V$, but only on the subset $E$ of edges of the mesh.  We also
note that $\varphi(x_0, x_1) = - \varphi(x_1, x_0)$ for any edge
$(x_0, x_1)$.

We continue by defining discrete two-forms as functions $\psi(x_0,
x_1, x_2, x_3)$ on the set of faces.  Again, these functions can be
extended unambiguously by linearity to a proper cochain.  To
summarize, we have the following definition:

\begin{definition}
  For $n = 0, 1, 2$, a \define{discrete $n$-form} is a linear map $f:
  C_n \rightarrow \bfR$.  For $n > 2$, all discrete $n$-forms are
  zero.  The set of all discrete $n$-forms is denoted by $C_n^\ast$. 
\end{definition}

For the sake of self-containedness, we recall the explicit form of the
coboundary operator $\d: C_n^\ast \rightarrow C_{n+1}^\ast$.  For
a zero-form $\phi$, $\d \phi(x_0, x_1) = \phi(x_1) - \phi(x_0)$.
For a one-form $\varphi$,
\begin{equation} \label{eqdiff} 
  \d \varphi(\gothf) = \varphi(x_0, x_1)
    + \varphi(x_1, x_2) + \varphi(x_2, x_3) + \varphi(x_3, x_0)
  \qforall \gothf = (x_0, x_1, x_2, x_3).
\end{equation}
The coboundary of a discrete two-form is defined to be zero.  We will
sometimes refer to $\d$ as the ``discrete differential''.  Note that
$\d$ is the dual of the boundary operator $\partial: C_{n + 1}
\rightarrow C_n$, in the sense that $\left< \d f, v \right> =
\left<f, \partial v\right>$ for all $f \in C^\ast_n$ and $v \in C_n$,
and where $\left< \cdot, \cdot \right>$ is the natural pairing between
$C_n$ and $C^\ast_n$.  This is a discrete analogue of the Stokes
theorem.  For more information, see~\cite[p.~186]{Hatcher}.

\subsection{The discrete Hodge star} \label{sec:hodge}

In Riemannian geometry, the Hodge star on an oriented manifold $M$
maps $n$-forms on $M$ to $(m - n)$-forms on $M$, where $m = \dim M$.
As could be expected from the continuous theory, the discrete Hodge
star $\star$, to be introduced below, maps discrete $n$-forms into
$(2-n)$-forms.  However, there is an additional complication in the
discrete case: the forms $\star f$ are not defined on the mesh itself,
but rather on a \emph{dual mesh}, which we now define.

\begin{figure}
\begin{center}
  \includegraphics[scale=0.7]{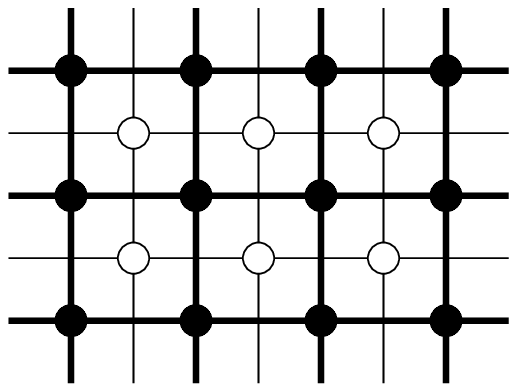}
  \caption{Square mesh (black) and its dual (light).}
  \label{fig:vierkantjes}
\end{center}
\end{figure}

The \define{dual mesh} $(V^\ast, E^\ast)$ is constructed as follows.
For every face in $F$, there is a vertex in $V^\ast$ (for which one
usually takes the circumcentric dual; see \cite{DEC}).  There is an
edge in $E^\ast$ between two vertices $q_0, q_1 \in V^\ast$ if and
only if the faces in $F$ corresponding to $q_0$ and $q_1$ have
precisely one edge in common.  This determines the sets $V^\ast$ and
$E^\ast$; the set $F^\ast$ consists of the faces of this dual graph.
It is easy to see that to each face in $F^\ast$, there corresponds a
vertex in $V$.  See figure~\ref{fig:vierkantjes} for an illustration.

Implicit in this definition is the existence of a duality operator
$\ast$ between $n$-chains on the mesh and $(2-n)$-chains on the dual
mesh.  This duality is well defined, but only up to orientation; we
now use the standard orientation of $\bfR^2$ to settle this point.
The definition used here agrees with the algorithm for the orientation
of dual cells proposed in \cite[remark~2.5.1]{HiraniPhd}.

We first define $\ast$ on the elements of $V$, $E$, and $F$.  By
linear extension, it will then be determined on the whole of $C_n$.
The dual vertex $\ast \gothf$ of a face $\gothf$ whose vertices are
written down in anticlockwise fashion is the element $r$ of $V^\ast$
introduced above.  If, on the other hand, the orientation of $\gothf$
is the opposite, then $\ast \gothf$ equals $-r$.  We define the dual
$\ast x$ of a vertex $x$ to be the corresponding face in $F^\ast$,
with the natural orientation: $\ast x = (r_1, \ldots, r_4)$.  Finally,
the definition of $\ast$ on $E$ is slightly more intricate; here, we
follow \cite{HiraniPhd}.  Let $(x_0, x_1)$ be an edge in $E$ and let
$\{r_0, r_1\}$ be the corresponding dual edge, considered as an
unordered set.  The line segments $[x_0, x_1]$ and $[r_0, r_1]$
determine a basis of $\bfR^2$: if this basis is positively oriented,
then $\ast(x_0, x_1) = (r_0, r_1)$, otherwise, $\ast(x_0, x_1) = (r_1,
r_0)$.  In the case of the square mesh of
figure~\ref{fig:vierkantjes}, the action of $\ast$ on $E$ corresponds
to an anticlockwise rotation over $\pi/2$.

On the dual mesh, one can again introduce discrete forms.  We will
denote the vector space of discrete $n$-forms on the dual mesh by
$D^\ast_n$.  As the dual mesh is again a square mesh, there is a
natural way to extend $\ast$ to an operator from $(V^\ast, E^\ast)$ to
$(V, E)$.  It is then easy to check that $\ast \ast v = (-1)^{n(2-n)}
v$ for any $v \in C_n$.

\begin{definition}
  The \define{discrete Hodge star} $\star: C^\ast_n \rightarrow
  D^\ast_{2-n}$ is defined by 
  \[ 
    (\star \alpha)(\ast v) = \alpha(v) \qforall\, v \in C_n.
  \]
\end{definition}

The definition given here is (up to a constant) a special case of the
one proposed in \cite{DEC}.  Note that $\star \star \alpha =
(-1)^{n(2-n)}\alpha$.

With the discrete Hodge star and the coboundary operator of the
previous paragraph, we now arrive at the definition of the discrete
codifferential.

\begin{definition}
  Let $\alpha$ be a discrete $n$-form.  Then the \define{discrete
    codifferential} $\delta \alpha$ is the discrete $(n-1)$-form
  $\delta \alpha$ defined as $\delta \alpha = \star\, \d \star
  \alpha$.
\end{definition}

It is useful to write out this definition explicitly for one-forms
and two-forms.  If $\varphi$ is a discrete one-form, then $\delta
\varphi$ is given by
\[
  (\delta \varphi)(x) = \varphi(x_1, x) + \varphi(x_2, x) +
  \varphi(x_3, x) + \varphi(x_4, x),
\]
where $x_1, x_2, x_3, x_4$ are the end points of the edges that
emanate from $x$.  In other words, $\delta \varphi$ assigns to each
vertex $x$ the sum of contributions from $\varphi$ on the edges that
have $x$ as a vertex.  Secondly, for a discrete two-form $\psi$ we
note that $\delta \psi$ is the discrete one-form given by $\delta \psi
(x_0, x_1) = \psi(\gothf_0) - \psi(\gothf_1)$, where $\gothf_0$ and
$\gothf_1$ are the faces that have $(x_0, x_1)$ as a common edge, and
where $\gothf_0$ is the face where the orientation of the boundary
edges agrees with the ordering of $(x_0, x_1)$, whereas $\gothf_1$ is
the face with the opposite ordering.

\begin{remark}
  Our definition of discrete codifferential agrees with the one from
  Desbrun \textit{et al.}  \cite{DEC} up to sign.  In their paper, the
  discrete codifferential is defined as $\delta \alpha= (-1)^{kn + 1}
  \star\, \d \star \alpha$, where $k$ is the dimension of the ambient
  space.  Note that here, as $k = 2$, $(-1)^{kn + 1} = -1$ regardless
  of $n$.
\end{remark}

\subsection{Discrete connections}

In the preceding sections, we introduced discrete one-forms as
assignments of a real number to each edge $\gothe \in E$.  This theory
can be extended in a straightforward way to discrete forms taking
values in an arbitrary \emph{Abelian} Lie group $G$, the only
significant difference being that we have to redefine the spaces of
$n$-chains $C_n$ as consisting of formal linear combinations with
coefficients in $\bfZ$.  

For instance, if $\varphi: C_1 \rightarrow G$ is a discrete one-form,
then $\d\varphi$ is determined by its action on the set $F$ by
(\ref{eqdiff}) and can be extended by linearity to yield a map from
$C_2$ to $G$, where it should be borne in mind that the elements of
$C_2$ are still formal linear combinations of elements in $F$, but now
with coefficients in $\bfZ$.

The theory of discrete forms with values in an Abelian Lie group will
be used in section~\ref{sec:MV}, but in the general case, we will be
confronted with mappings from $E$ to a \emph{non-Abelian} Lie group
$G$.  Such maps can no longer be interpreted as discrete one-forms.
Luckily, it turns out that these maps have a natural interpretation as
\emph{discrete $G$-connections}, which we now define.

\begin{definition} \label{def:DC} A \define{discrete $G$-connection}
  is a map $\omega: E \rightarrow G$, such that, for all edges $\gothe
  \in E$, $\omega(\gothe^{-1}) = \omega(\gothe)^{-1}$.  The
  \define{curvature} of such a connection is the map $\Omega: F
  \rightarrow G$ defined as $\Omega(\gothf) = \omega(\gothe_1) \cdots
  \omega(\gothe_4)$, where $\gothe_1, \ldots, \gothe_4$ are the
  boundary edges of the face $\gothf$.  A discrete $G$-connection is
  said to be \define{flat} if $\Omega(\gothf) = e$ for all
  $\gothf \in F$.
\end{definition}

Note that in the case of a non-flat connection, $\Omega(\gothf)$
depends not only on $\gothf$, but also on the exact representation of
$\gothf$ as a set of edges $\gothe_1, \ldots, \gothe_4$ (any cyclic
permutation of this set represents the same face).  However, this
indeterminacy does not occur for flat connections, the only case that
we will consider later on.

The theory of discrete $G$-connections closely mimics the usual
theory of connections. As an example, we mention the following
proposition, from which a number of interesting properties may be
deduced. 

\begin{prop} \label{prop:holonomy} Consider a discrete
  $G$-connection $\omega: E \rightarrow G$.  If $\omega$ is flat, then
  there exists a unique mapping $\phi: V \times V \rightarrow G$ such
  that $\phi_{|E} = \omega$.  
\end{prop}
This follows immediately from \cite[prop.~7]{discreet}, or can be
proved directly as follows.

We define a \emph{path} in $E$ to be a sequence $\gothe_1, \gothe_2,
\ldots, \gothe_m$ of edges, such that the end vertex of $\gothe_i$ is
the begin vertex of $\gothe_{i+1}$ (for $i = 1, \ldots, m-1$).  Let
$\omega: E \rightarrow G$ be any discrete $G$-connection. In
particular, $\omega$ need not be flat.  Then, $\omega$ induces a map
$\hat{\omega}$ from the set of paths to $G$ as follows:
\[
\hat{\omega}(\gothe_1, \gothe_2, \ldots, \gothe_m) =
\omega(\gothe_1)\omega(\gothe_2) \cdots \omega(\gothe_m).
\]

We call $\hat{\omega}$ the ``discrete holonomy mapping''.  Note that
if $\omega$ is a flat connection, one can easily prove that
$\hat{\omega}$ maps closed paths to the unit in $G$.  Furthermore, in
the case of a flat connection, it can be easily established that
simplicially homotopic paths have the same image under $\hat{\omega}$.
(This is the discrete counterpart of a well-known theorem of continuous
connections (see \cite[p.~93]{KN1}): if $\omega$ is a flat connection,
then any two closed homotopic loops have the same holonomy.)

Let us now return to the discrete theory.  In the case of a flat
connection on a simply-connected cell complex, we have seen that
$\hat{\omega}$ does not depend on the choice of path between two fixed
points, and induces therefore a map $\hat{\omega}: V \times V
\rightarrow G$.  This map satisfies the requirements in
proposition~\ref{prop:holonomy}.  

\begin{remark}
  The concept of discrete $G$-connections used here is common in
  lattice gauge theories (see \cite{Wise}).  Novikov \cite{Novikov}
  studied a similar concept.  

  Closely related to this definition is the concept of \emph{discrete
    connection on a discrete principal fibre bundle} (see
  \cite{leokconn} for definitions and \cite[\S5.4]{GroupoidMech05} for
  an application to discrete reduction).  The latter could be used to
  extend the results in this paper to the case of
  \emph{Lagrange-Poincar\'e reduction}, where a $G$-invariant
  Lagrangian on a pair groupoid $Q \times Q$ is given.  If $Q$
  coincides with the symmetry group $G$, then Lagrange-Poincar\'e
  reduction is just Euler-Poincar\'e reduction, the case considered
  here.
\end{remark}

\section{Discrete Euler-Poincar\'e reduction} \label{sec:DEP}

In this section, we begin our study of discrete Lagrangian field
theories.  Initially, we define discrete fields as follows:
\begin{definition}
  A \emph{discrete field} is a map $\phi: V \rightarrow G$.
\end{definition}
If the discrete Lagrangian, to be defined below, is $G$-invariant, we
shall see that these fields induce a new class of discrete fields,
that associate a group element to each edge.  From the last section,
we know that such maps have a natural interpretation as discrete
$G$-connections.

In theorem~\ref{thm:eulerpoincare}, it is shown how the field
equations for the unreduced fields $\phi$ are equivalent to a set of
equations, called \emph{discrete Euler-Poincar\'e equations}, for the
reduced fields $\varphi$.  Both sets of equations arise by extremizing
a certain action functional.  In theorem~\ref{thm:reconstruction}, we
deal with the reconstruction problem.  Starting from a reduced field
$\varphi: E \rightarrow G$, it is shown that $\varphi$ gives rise to a
solution $\phi: V \rightarrow G$ of the original field equations if
and only if the curvature of $\varphi$ vanishes.  This treatment was
inspired by the work of Castrill\'on and Ratiu \cite{MarcoRatiu}, who
developed Lagrangian reduction for field theories in the continuous
case.

% \begin{remark}
%   From now on, we restrict ourselves to the square mesh defined in
%   remark~\ref{rem:square}.
% \end{remark}

\subsection{The discrete Lagrangian} \label{sec:disclagrangian}

\begin{definition}
  A \define{discrete Lagrangian} is a function $L : G \times G \times G
  \rightarrow \bfR$.
\end{definition}

A word of explanation is in order here.  In the continuous case, a
Lagrangian is a function $\calL : J^1 \pi \rightarrow \bfR$, where
$J^1\pi$ is the first jet bundle of the bundle $\pi: \bfR^2 \times G
\rightarrow \bfR^2$.  It can be shown that, in this case, $J^1\pi$ is
isomorphic to $\bfR^2 \times (TG \oplus TG)$ (see
\cite[lemma~4.1.20]{Saunders89}).  Under the (modest) assumption that
Lagrangian $\calL$ does not depend on the coordinates of the base
space, we therefore conclude that $\calL$ is just a function on $TG
\oplus TG$.

\begin{figure}
\begin{center}
  \includegraphics[scale=0.7]{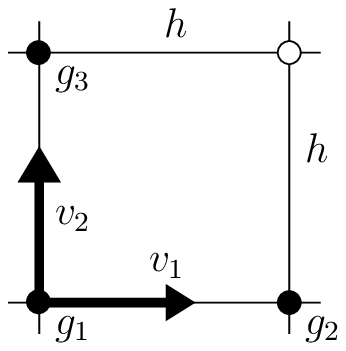}
  \caption{Discretization of $TG \oplus TG$ by $G \times G \times G$}
  \label{fig:points}
\end{center}
\end{figure}

The idea of defining a discrete Lagrangian as a function on $G \times
G \times G$ then follows from the idea of Moser and Veselov of
approximating $TG$ by $G \times G$: by applying this Veselov-type
discretization twice (once in the ``horizontal'' and once in the
``vertical'' direction of the square mesh), it follows that $TG \oplus
TG$ has $G \times G \times G$ as a natural discrete counterpart; see
figure~\ref{fig:points}.

The Lie group $G$ has a natural diagonal action by left translations
on the Cartesian product $G \times G$: $g \cdot (g_1, g_2) = (gg_1,
gg_2)$.  We now wish to study the situation where $L$ is invariant
under this action.  As the quotient $(G \times G)/G$ is naturally
isomorphic to the Lie group $G$ itself, a discrete field $\phi: V
\rightarrow G$ induces a ``reduced field'' $\varphi: E \rightarrow G$
as follows:
\[
  \varphi(\gothe) = \phi(x_0)^{-1} \phi(x_1), \qforall \gothe = (x_0,
  x_1) \in E.
\]
This leads us to the following definition:
\begin{definition}
  A \emph{reduced discrete field} is a map $\varphi: E \rightarrow
  G$ such that $\varphi(\gothe^{-1}) = \varphi(\gothe)^{-1}$.
  Equivalently, it is a discrete $G$-connection.
\end{definition}

The reduced field $\varphi$ induced by a discrete field $\phi: V
\rightarrow G$ is flat when considered as a discrete $G$-connection.
Hence, by proposition~\ref{prop:holonomy} (or simply by construction),
it may be extended to a map $\varphi: V \times V \rightarrow G$.
However, not all reduced fields are flat.

The Lagrangian ${L}$ gives rise to a reduced Lagrangian
$l: G \times G \rightarrow \bfR$ defined by
\[
  l(g_1^{-1}g_2, g_1^{-1}g_3) = L(g_1, g_2, g_3).
\]

Let $\Phi: G \times G \rightarrow G$ be the map defined as $\Phi(g,
g') = g^{-1}g'$, so that the reduced field $\varphi$ associated to a
discrete field $\phi$ is given by
\begin{equation} \label{correspondence}
  \varphi = \Phi \circ (\phi \times \phi).
\end{equation}
Furthermore, we introduce the following
mapping: 
\begin{equation} \label{mapbottom}
  \hat{\Psi}: (g_1, g_2, g_3) \in G \times G \times G \mapsto
  (g_1^{-1}g_2, g_1^{-1}g_3) \in G \times G.
\end{equation}
It is obvious that the Lagrangian $L$ is related to the reduced
Lagrangian by $L = l \circ \hat{\Psi}$.  For future reference, we also
introduce a map $\bar{\Psi}: T(G^3) \rightarrow (G \times \gothg)^3$
as
\begin{equation} \label{maptop} \bar{\Psi}(v_1, v_2, v_3) = ( g_1,
  g_2, g_3; TL_{g_1^{-1}}(v_1), TL_{g_2^{-1}}(v_2),
  TL_{g_3^{-1}}(v_3)),
\end{equation}
where $v_i \in T_{g_i} G$.

%\begin{remark}[Notation]
An unreduced field $\phi : V \rightarrow G$ can be interpreted as an
assignment of an element $\phi_{i, j}$ in $G$ to each vertex $(i, j)$.
Similarly, a reduced field $\varphi: E \rightarrow G$ can be described
as an assignment of a group element $u_{i,j}$ to each ``vertical''
edge $((i,j+1),(i,j))$, and of a group element $v_{i,j}$ to each
``horizontal'' edge $((i,j), (i+1,j))$, where
\begin{equation}
  u_{i,j} = \phi_{i,j}^{-1}\phi_{i+1,j}  \qqand v_{i,j} = \phi_{i,j}^{-1}
  \phi_{i, j+1}. \label{eq:reduced}
\end{equation}
We will use these notations for the remainder of the paper.
%\end{remark}

\subsection{The reduction problem} \label{sec:redproblem}

Given a discrete Lagrangian $L$, the discrete action sum $S$ is given
by 
\[
  S(\phi) = \sum_{(i,j) \in U} L( \phi_{i,j}, \phi_{i+1, j}, \phi_{i,j+1}),
\]
where $\phi$ is a map from $V$ to $G$, and where $U$ is a finite
subset of the set of vertices $V$.  In \cite{MPS98}, it was shown that
$\phi$ is an extremum of this action if and only $\phi$ satisfies the
following set of discrete Euler-Lagrange equations: for all $(i, j)
\in V$, 
\begin{equation} \label{eq:EL}
  \frac{\partial L}{\partial g_1}(\phi_{i,j}, \phi_{i+1, j},
  \phi_{i,j+1}) + \frac{\partial L}{\partial g_2}(\phi_{i-1,j},
  \phi_{i,j}, \phi_{i-1,j+1}) + \frac{\partial L}{\partial
    g_3}(\phi_{i,j-1}, \phi_{i+1,j-1}, \phi_{i,j}) = 0.
\end{equation}
Similarly, we may define the reduced action sum $s$ as
\[
  s(\varphi) = \sum_{(i,j)\in U} l(u_{i,j}, v_{i,j}).
\]
A reduced field $\varphi: E \rightarrow G$ is an extremum of $s$ if
and only if it satisfies the \emph{discrete Euler-Poincar\'e}
equations, to be derived below.  The central aspects of discrete
Euler-Poincar\'e reduction are summarized in the following theorem.
This theorem, as well as its proof, are very similar to the discrete
reduction process in mechanics (see \cite{MarsdenPekarskyShkoller99}).

\begin{theorem}[Reduction] \label{thm:eulerpoincare} Let $L$ be a
  $G$-invariant Lagrangian on $G \times G \times G$ and consider the
  reduced Lagrangian $l$ on $G \times G$.  Consider a discrete field
  $\phi: V \rightarrow G$ and let $\varphi: V \times V \rightarrow G$
  be the induced reduced field.  Then the following are equivalent:
  \begin{AlphaList}
    \item $\phi$ is a solution of the discrete Euler-Lagrange
      equations for $L$; \label{thEL}
    \item $\phi$ is an extremum of the action sum $S$ for arbitrary
      variations; \label{thAS}
    \item the reduced field $\varphi$ is a solution of the discrete
      Euler-Poincar\'e equations: \label{thEP}
      \begin{equation} \label{eq:EP}
      \begin{split}
        \left[ \left( R_{u_{i,j}}^\ast \d l(\cdot, v_{i,j}) \right)_e
          - \left( L_{u_{i-1,j}}^\ast \d l(\cdot, v_{i-1,j}) \right)_e
        \right] & + \\
        \left[ \left( R_{v_{i,j}}^\ast \d l(u_{i,j}, \cdot) \right)_e
          - \left( L_{v_{i,j-1}}^\ast \d l(u_{i,j-1}, \cdot) \right)_e
        \right] & = 0;
      \end{split}
      \end{equation}
    \item the reduced field $\varphi$ is an extremum of the reduced
      action sum $s$ for variations of the form
      \begin{equation} \label{varU}
        \delta u_{i,j} = TL_{u_{i,j}}(\theta_{i+1,j}) -
        TR_{u_{i,j}}(\theta_{i,j}) \in T_{u_{i,j}} G 
      \end{equation}
      and
      \begin{equation} \label{varV} \delta v_{i,j} =
        TL_{v_{i,j}}(\theta_{i,j+1}) - TR_{v_{i,j}}(\theta_{i,j}) \in
        T_{v_{i,j}} G,
      \end{equation}
      where $\theta_{i,j} = TL_{\phi_{i,j}^{-1}}(\delta \phi_{i,j})
      \in \gothg$. \label{thRAS}
  \end{AlphaList}
\end{theorem}
\begin{proof}
The equivalence of (\ref{thEL}) and (\ref{thAS}) follows from a
standard argument in discrete Lagrangian field theories, and was
shown in \cite{discreet}.

In order to prove the equivalence of (\ref{thAS}) and (\ref{thRAS}),
we note that $L = l \circ \hat{\Psi}$ (see (\ref{mapbottom})), from
which we conclude that if $\varphi = \Phi \circ \phi$, then $S(\phi) =
s(\varphi)$.  Now, consider the components $\{u_{i,j}\}$ and
$\{v_{i,j}\}$ of the reduced field, as in (\ref{eq:reduced}).  It is
easy to check that an arbitrary variation $\epsilon \mapsto
\phi_{i,j}(\epsilon)$ of $\phi$ induces corresponding variations
$\delta u_{i,j}$ and $\delta v_{i,j}$ of $u_{i,j}$ and $v_{i,j}$,
given by (\ref{varU}) and (\ref{varV}).

Finally, we show the equivalence of (\ref{thRAS}) with the
Euler-Poincar\'e equations (\ref{eq:EP}).  From the reduced action sum
$s(\varphi)$ we obtain that
\begin{align*}
  \dde s(\varphi(\epsilon)) \Big|_{\epsilon=0} & = \sum_{i,j}
  \dde l(u_{i,j}(\epsilon),
  v_{i,j}(\epsilon)) \Big|_{\epsilon = 0} \\
  & = \sum_{i,j} \left( \d l(\cdot, v_{i,j}) \cdot \delta u_{i,j} + \d
    l(u_{i,j}, \cdot) \cdot \delta v_{i,j} \right).
\end{align*}
Substitution of (\ref{varU}) and (\ref{varV}) into this expression
then yields (after relabelling some of the summation indices) the
discrete Euler-Poincar\'e equations (\ref{eq:EP}).
\end{proof}

In this context, a variation can be interpreted in two ways.  In the
case of unreduced fields, a variation of a field $\phi_{i, j}$ is a
map $\delta \phi : V \rightarrow TG$ such that $\delta\phi_{i, j} \in
T_{\phi_{i, j}} G$.  In the case of reduced fields, a variation of a
field $\{u_{i, j}, v_{i, j}\}$ is a pair of maps $\delta u, \delta v:
V \rightarrow TG$ such that $\delta u_{i, j} \in T_{u_{i,j}} G$ and
$\delta v_{i,j} \in T_{v_{i,j}} G$.  In both cases, we demand that the
variation is zero on the boundary of $U$:
\begin{equation} \label{varzero}
  \delta \phi_{i,j} = 0, \qqand \delta u_{i,j} = \delta{v_{i, j}} = 0
  \qforall (i, j) \in \partial U.
\end{equation}
Here, the boundary $\partial U$ is defined as the set of vertices $(i,
j)$ such that $(i, j)$ is a vertex of at least one face in $U$, and at
least one face not in $U$ (see \cite{MPS98}).

\subsection{The reconstruction problem} \label{sec:reconstruction}

From theorem~\ref{thm:eulerpoincare}, we know that a solution $\phi :
V \rightarrow G$ of the discrete Euler-Lagrange equations gives rise
to a reduced field $\varphi : E \rightarrow G$, which has a natural
interpretation as a flat discrete connection in the sense of
definition~\ref{def:DC}.

To tackle the converse problem, we use the following consequence of
proposition~\ref{prop:holonomy}.  Recall that the holonomy mapping
$\hat{\omega}$ of a flat discrete connection $\omega$ is a map from $V
\times V$ to $G$.

\begin{prop} \label{prop:integrate} Let $\omega$ be a flat
  discrete $G$-connection with associated discrete holonomy
  $\hat{\omega}: V \times V \rightarrow G$.  Then there exists a map
  $\phi: V \rightarrow G$ such that $\omega(x_0, x_1) = \phi(x_0)^{-1}
  \phi(x_1)$.  The map $\phi$ is unique up to left translation by an
  element of $G$.
\end{prop}
\begin{proof}
  Choose an arbitrary vertex $x_0$ and a group element $g_0$, and
  define $\phi(x_0) = g_0$.  Let $x_1$ be any other vertex and put
  $\phi(x_1) = g_0 \hat{\omega}(x_0, x_1)$.  This map is well
  defined. 
\end{proof}

Let $\varphi: E \rightarrow G$ be a solution of the discrete
Euler-Poincar\'e equations (\ref{eq:EP}).  We now wish to reconstruct
a solution $\phi$ of the original problem, such that $\varphi = \Phi
\circ \phi$.  The map $\phi$ is provided by
proposition~\ref{prop:integrate}, on the condition that $\varphi$ is a
flat connection.  As soon as $\varphi$ is not flat, the holonomy
$\hat{\omega}$ is path dependent, and no such $\phi$ can exist.
Therefore, we have the following theorem.

\begin{theorem}[reconstruction] \label{thm:reconstruction} Let
  $\varphi: E \rightarrow G$ be a solution of the discrete
  Euler-Poincar\'e equations (\ref{eq:EP}).  There exists a solution
  $\phi: V \rightarrow G$ of the unreduced Euler-Lagrange equations
  (\ref{eq:EL}) if and only if $\varphi$ is flat.  In that case,
  $\phi$ is uniquely determined up to left translation by an element
  of $G$.
\end{theorem}

\begin{remark}
  In some cases, the element $g_0 \in G$ used in constructing the map
  $\phi$ is fixed by considering boundary or initial
  conditions on the solutions.
\end{remark}

\section{The Noether theorem} \label{sec:noether}

In the case of continuous field theories, Noether's theorem states
that, for each continuous symmetry, there exists a conservation law
(see \cite{symm04} for an overview).  Here, we will show that a
similar theorem holds for discrete field theories with a continuous
symmetry, and that the conservation law associated to the left
$G$-invariance of the Lagrangian is equivalent to the Euler-Poincar\'e
equations.  A similar theorem was proved in \cite{MarcoLMP} in the
case of Lagrangian reduction for continuous field theories.

\subsection{The Poincar\'e-Cartan forms} \label{sec:PC}

We begin by introducing a set of Poincar\'e-Cartan forms that will be
instrumental in formulating the Noether theorem.  It should be
remarked that the definition of the Poincar\'e-Cartan forms here is
a special case of a more general construction, elaborated in
\cite{discreet} and briefly outlined in section~\ref{sec:general}.

As in section~{\ref{sec:DEP}}, let $L : G^3 = G \times G \times G
\rightarrow \bfR$ be a $G$-invariant Lagrangian and consider the
reduced Lagrangian $l: G^2 = G \times G \rightarrow \bfR$.  Associated
to $L$ is a set of three one-forms, called \define{Poincar\'e-Cartan
  forms}, and defined as follows:
\[
\theta_{(1)}^L : G^3 \rightarrow T^\ast (G^3), \quad
\dual{\theta_{(1)}^L(g_1, g_2, g_3), (v_1, v_2, v_3)} = \dual{\d
  L(\cdot, g_2, g_3)_{g_1}, v_1},
\]
with $v_i \in T_{g_i} G$, and similarly for $\theta_{(2)}^L$ and
$\theta_{(3)}^L$.  Note that $\theta_{(1)}^L + \theta_{(2)}^L +
\theta_{(3)}^L = \d L$.

For the reduced Lagrangian $l$, the definition of the
Poincar\'e-Cartan forms is somewhat less direct.  We put
\begin{gather} \theta_{(1)}^l:  G^2 \rightarrow
  \gothg^\ast \oplus \gothg^\ast \oplus \gothg^\ast, \label{def1} \\
  \dual{\theta_{(1)}^l(u, v),(\xi_1, \xi_2, \xi_3)} = - \dual{\d l(\cdot,
    v), TR_{u}(\xi_1)} - \dual{\d l(u, \cdot), TR_{v}(\xi_1)},
  \nonumber 
\end{gather}
where $\xi_1, \xi_2, \xi_3 \in \gothg$.  Alternatively, if $t \mapsto
h(t)$ is a curve in $G$ such that $h(0) = e$ and $\dot{h}(0) = \xi_1$,
$\theta_{(1)}^l$ may be defined as
\[
\dual{\theta_{(1)}^l(u,v),(\xi_1, \xi_2, \xi_3)} = \ddt l(h(t)^{-1}u,
h(t)^{-1}v) \Big|_{t = 0}.
\]
The remaining one-forms $\theta_{(2)}^l$ and $\theta_{(3)}^l$ are
defined by
\begin{equation} \label{def2} \dual{\theta_{(2)}^l(u,v),(\xi_1,
    \xi_2, \xi_3)} = \dual{\d l(\cdot, v), TL_{u}(\xi_2)}
\end{equation}
and
\begin{equation} \label{def3} \dual{\theta_{(3)}^l(u,v), (\xi_1,
    \xi_2, \xi_3)} = \dual{\d l(u, \cdot), TL_{v}(\xi_3)}.
\end{equation}
The apparent asymmetry between (\ref{def1}) and (\ref{def2},
\ref{def3}) is due to our definition of $l$.  For a fully symmetric
set of Poincar\'e-Cartan forms, one should follow the prescriptions of
\cite{discreet}.

The Lagrangian $L$ is related to the reduced Lagrangian $l$ as
follows: $L = l \circ \hat{\Psi}$.  A straightforward computation (or
the application of theorem~25 in \cite{discreet}) shows us that the
corresponding Poincar\'e-Cartan forms are related in a similar way:

\begin{lemma} \label{lemma:pullback}
  Let $\Psi$ be the bundle map $(\hat{\Psi}, \bar{\Psi})$ as defined
  above.  Then 
  \[
  \Psi^\star \theta^l_{(i)} = \theta^L_{(i)}.
  \]
  Furthermore, a similar identity holds for the presymplectic forms
  $\Omega^l_{(i)} := - \d \theta^l_{(i)}$ and $\Omega^L_{(i)} := - \d
  \theta^L_{(i)}$: 
  \[
    \Psi^\star \Omega^l_{(i)} = \Omega^L_{(i)}.
  \]
\end{lemma}

We recall that the pullback of $\theta_{(i)}^l$ by the bundle map
$\Psi = (\hat{\Psi}, \bar{\Psi}): T(G)^3 \rightarrow (G \times
\gothg)^3$ is defined as follows:
\[
\dual{(\Psi^\star \theta_{(i)}^l)(g_1, g_2, g_3),(v_1, v_2, v_3)} =
\dual{\theta_{(i)}^l(\hat{\Psi}(g_1, g_2, g_3)), \bar{\Psi}(v_1, v_2,
  v_3)}.
\]

% \begin{remark}
%   The authors of \cite{MPS98} originally used a different method to
%   derive the Poincar\'e-Cartan forms, which stresses the relation
%   between the Poincar\'e-Cartan forms and the variational background
%   of the discrete Euler-Lagrange equations.  They start from the space
%   of all fields $\phi: V \rightarrow G$.  A tangent vector at a field
%   $\phi$ to this space is then a variation in the sense of
%   section~\ref{sec:redproblem}, but not necessarily vanishing on the
%   boundary of the domain of definition of $\phi$.  However, one can
%   express the exterior differential of $S$ at $\phi$ as follows:
%   \begin{multline} \label{varformula} \d S_\phi( \delta \phi) =
%     \sum_{(i, j) \in \interior{U}} \left( (\text{E.L.})_{i,j} \right)
%     \cdot \delta\phi_{i,j} \\ + \sum_{(i,j) \in \partial U} \left(
%       \sum_{\gothf: \gothf\in U, (i, j) = \gothf_l}
%       \theta^L_{(l)}(\psi(\gothf))(\delta \phi(\gothf_1), \delta
%       \phi(\gothf_2), \delta \phi(\gothf_3)) \right),
%   \end{multline}
%   where $(\text{E.L.})_{i,j}$ is the left-hand side of (\ref{eq:EL}),
%   and where the summation in the second term is over all faces $\gothf
%   \in U$ such that $(i, j)$ is a vertex of $\gothf$.  Furthermore,
%   $\psi$ is the map defined as $\psi = \phi \times \phi \times \phi$.
%   The interior $\interior{U}$ of $U$ is defined as $\interior{U} = U
%   - \partial U$.  A similar formula as (\ref{varformula}) can be
%   proved for the reduced action $s$.
% \end{remark}

\subsection{The unreduced Lagrangian}

The Lagrangian $L : G^3 \rightarrow \bfR$ is assumed to be left
$G$-invariant in the sense that $L(gg_1, gg_2, gg_3) = L(g_1, g_2,
g_3)$ for all $g$ in $G$.  According to Noether's theorem, which we
will prove in a moment, there is a conservation law associated to this
symmetry.

Let $\xi$ be an element of $\gothg$.  Infinitesimal invariance of
Lagrangian under the flow generated by $\xi$ is expressed as
\begin{equation} \label{invariance}
  \dual{\d L(g_1, g_2, g_3),  (\xi_G(g_1), \xi_G(g_2), \xi_G(g_3))} =
  0,
\end{equation}
where $\xi_G$, defined by $\xi_G(g) = TR_g(\xi)$, is the fundamental
vector field associated to $\xi$.

Consider the functions $J^i_\xi: G^3 \rightarrow \bfR$, $i=1, 2, 3$,
given by
\[
  J^i_\xi(g_1, g_2, g_3) = \dual{\theta^L_{(i)}(g_1, g_2, g_3),
    (\xi_G(g_1), \xi_G(g_2), \xi_G(g_3))}.
\]
As $J^i_\xi$ is linear in $\xi$, we can define a map $J^i: G^3
\rightarrow \gothg^\ast$ by the prescription $\dual{J^i, \xi} =
J^i_\xi$, for all $\xi \in \gothg$.  An immediate consequence of the
$G$-invariance is that
\begin{equation} \label{redundancy}
J^1 + J^2 + J^3 = 0.
\end{equation}

Now, let there be given a map $\phi: V \rightarrow G$ and consider the
pull-back of each $J^i$ by $\psi := \phi \times \phi \times \phi$.  In
this way, we obtain a map from $F$ to $\gothg^\ast$, which can be
identified, by means of the discrete Hodge star, with a map from
$V^\ast$ to $\gothg^\ast$.  In view of (\ref{redundancy}) it turns out
that we only need to consider two such maps, which we will denote by
$\eta^{(x)}$ and $\eta^{(y)}$, and which are given by 
\[
  \eta^{(x)}(r) = J^2(\phi(x_1), \phi(x_2), \phi(x_3)), 
  \qqand
  \eta^{(y)}(r) = J^3(\phi(x_1), \phi(x_2), \phi(x_3))
\]
for $r \in V^\ast$, and where $\gothf = (x_1, x_2, x_3, x_4)$ is the
face dual to $r$: $\ast \gothf = r$.  We further put $\eta^{(x)}_\xi =
\dual{\eta^{(x)}, \xi}$ and $\eta^{(y)}_\xi = \dual{\eta^{(y)},
  \xi}$. 

\begin{figure}
\begin{center}
  \includegraphics[scale=0.7]{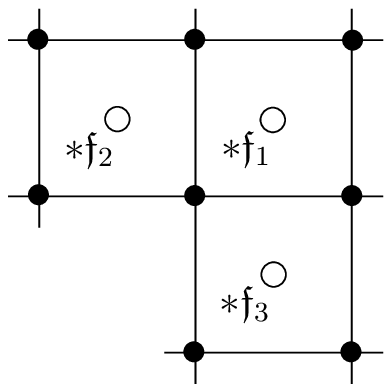}
  \caption{Location of $\gothf_1$, $\gothf_2$, and $\gothf_3$.}
  \label{fig:three}
\end{center}
\end{figure}

Before stating Noether's theorem, we recall the definition of the
backward difference operators $\delta^-_x$ and $\delta^-_y$: if $r_1$,
$r_2$ and $r_3$, with $r_i = \ast \gothf_i$, are located as in
figure~\ref{fig:three}, and $f: V^\ast \rightarrow \bfR$ is a
function, then
\[
(\delta^-_x f)(r_1) = f(r_1) - f(r_2) \qqand (\delta^-_y f)(r_1) =
f(r_1) - f(r_3).
\]
These difference operators can be extended without difficulty to the
case of vector-valued functions.

\begin{prop}[Noether] \label{prop:noether} Consider a
  $G$-invariant Lagrangian $L : G^3 \rightarrow \bfR$.  If $\phi$ is a
  solution of the discrete field equations (\ref{eq:EL}), then the
  maps $\eta^{(x)}, \eta^{(y)}: V^\ast \rightarrow \gothg^\ast$
  satisfy the following conservation law:
  \begin{equation} \label{noether}
  \delta^-_x \eta^{(x)} + \delta^-_y \eta^{(y)} = 0.
  \end{equation}
  Conversely, if $\phi$ is such that the associated mappings
  $\eta^{(x)}$ and $\eta^{(y)}$ satisfy (\ref{noether}), then $\phi$
  is a solution of the discrete field equations.
\end{prop}
\begin{proof}
  Consider points $r_1, r_2, r_3$ of the dual mesh as in
  figure~\ref{fig:three} and let $\gothf_i$ be the dual face to
  $r_i$: $\gothf_i = \ast r_i$.  Then we have that
\begin{align}
  \delta^-_x \eta^{(x)}_\xi(r_1) + \delta^-_y \eta^{(y)}_\xi(r_1) & =  
    \eta^{(x)}_\xi(r_1) - \eta^{(x)}_\xi(r_2) + \eta^{(y)}_\xi(r_1) -
    \eta^{(y)}_\xi(r_3) \nonumber \\
    & = J^2_\xi(\psi(\gothf_1)) - J^2_\xi(\psi(\gothf_2)) +
    J^3_\xi(\psi(\gothf_1)) - J^3_\xi(\psi(\gothf_3)) \nonumber \\
    & = -J^1_\xi(\psi(\gothf_1)) - J^2_\xi(\psi(\gothf_2)) -
    J^3_\xi(\psi(\gothf_3)), \label{consrewrite}
\end{align}
where we have used (\ref{redundancy}).  Recalling that $\psi$ is
defined as $\psi(\gothf) = (\phi(x_1), \phi(x_2), \phi(x_3))$, we note
that 
\[
  J^i_\xi(\psi(\gothf)) = 
  \theta^L_{(i)}(\psi(\gothf))(\xi_G, \xi_G, \xi_G) = \dual{
    \frac{\partial L}{\partial g_i}(\psi(\gothf)), \xi_G}.
\]
Therefore, we can rewrite the conservation law (\ref{consrewrite}) as
\[
  \delta^-_x \eta^{(x)}_\xi(r_1) + \delta^-_y \eta^{(y)}_\xi(r_1) = 
  - \dual{  \frac{\partial L}{\partial g_1}(\psi(\gothf_1)) +
    \frac{\partial L}{\partial g_2}(\psi(\gothf_2)) + \frac{\partial
      L}{\partial g_3}(\psi(\gothf_3)), \xi_G}, 
\]
which vanishes if $\phi$ satisfies the field equations.  Conversely,
if the left hand side is zero, then $\phi$ is a solution of the field
equations. 
\end{proof}

% \begin{remark} \label{rem:noether} Proposition~\ref{prop:noether} can
%   be extended in many ways.  On the one hand, one can envisage a
%   generalization where the symmetry group is different from $G$.  On
%   the other hand, it is possible to weaken the invariance hypothesis
%   (\ref{invariance}) by including on the right-hand side a term which
%   makes no contribution to the discrete action sum.  Such a term can
%   be viewed as a ``discrete divergence''.
% \end{remark}

\subsection{The reduced Lagrangian}

Not only is Noether's theorem equivalent to the unreduced discrete
field equations, it will turn out that it contains the
Euler-Poincar\'e equations as well.  To show this, we start from the
discrete conservation law as expressed by (\ref{consrewrite}). We use
the same notational conventions as in the preceding section and write
\[
  \hat{\Psi}(\psi(\gothf_i)) = (u_i, v_i), \quad i = 1, 2, 3.
\]
Furthermore, we note that $(\psi(\gothf_1))_1 = (\psi(\gothf_2))_2 =
(\psi(\gothf_3))_3$; this unique element of $G$ is denoted by $g$.  By
rewriting each of the three expressions in (\ref{consrewrite}) in
terms of the reduced Lagrangian $l$ only, we obtain
\[
  J^1_\xi(\psi(\gothf_1)) = - \dual{R_{u_1}^\ast \d l(\cdot, v_1),
    \eta} - \dual{R_{v_1}^\ast \d l(u_1, \cdot), \eta},
\]
where $\eta = \Ad_{g^{-1}} \xi$, as well as
\[
  J^2_\xi(\psi(\gothf_2)) = \dual{L_{u_2}^\ast \d l(\cdot, v_2), \eta}
  \qqand J^3_\xi(\psi(\gothf_3)) = \dual{L_{v_3}^\ast \d l(u_3,
    \cdot), \eta}.
\]
Putting all of these expressions together gives the following:
\begin{multline}
\delta^-_x \eta^{(x)}_\xi(r_1) + \delta^-_y \eta^{(y)}_\xi(r_1) = 
   \Big<  
      \left[R_{u_1}^\ast \d l(\cdot, v_1) - L_{u_2}^\ast \d l(\cdot,
        v_2) \right] + \\
      \left[R_{v_1}^\ast \d l(u_1, \cdot) - L_{v_3}^\ast \d l(u_3,
        \cdot) \right], \eta \Big>.
\end{multline}
As $\eta$ ranges over the whole of $\gothg$, we conclude that the
conservation law (\ref{noether}) implies the discrete Euler-Poincar\'e
equation (\ref{eq:EP}).

\section{Extending the Moser-Veselov approach} \label{sec:MV}

In their seminal paper, Moser and Veselov \cite{MoserVeselov91}
approached the problem of finding an integrable discretization of the
rigid-body equations by embedding the group $SO(n)$ into a linear
space, namely $\gl(n)$.  Somewhat later, Marsden, Pekarsky, and
Shkoller \cite{MarsdenPekarskyShkoller99} then developed a general
procedure of Lagrangian reduction for discrete mechanical systems, and
showed that the Moser-Veselov equations are equivalent to the discrete
Lie-Poisson equations.

Here, we intend to do the same thing for a fundamental model in field
theory: that of harmonic mappings from $\bfR^2$ into a Lie group $G$.
We will show that it is possible to develop a Moser-Veselov type
discretization of these field equations, provided that $G$ is embedded
in a linear space.  As could be expected, these discrete field
equations are equivalent to the Euler-Poincar\'e equations.

In the continuous case, the harmonic mapping Lagrangian is given by
\begin{equation} \label{HMlag}
\calL = \frac{1}{2} \dual{\phi^{-1} \phi_x, \phi^{-1}\phi_x} +
\frac{1}{2} \dual{\phi^{-1} \phi_y, \phi^{-1}\phi_y},
\end{equation}
where $\dual{\cdot, \cdot}$ is the Killing form on $\gothg$ and where
the subscript `$x$' and `$y$' denote partial differentiation with
respect to that variable.  For the sake of clarity, we will only treat
the case of harmonic maps that take values in $SO(n)$, embedded in
$\gl(n)$, in which case the Killing form is just the trace.  We stress
that the entire theory can be generalized to the case of an arbitrary
semi-simple group $G$, embedded in a linear space.

Consider the rectangular lattice from section~\ref{sec:mesh} and denote
the lattice spacing by $h$.  As usual, we denote the values of the
field $\phi$ on the lattice points by $\phi_{i,j}$.  We discretize the
reduced partial derivatives $\phi^{-1} \phi_x$ and $\phi^{-1}\phi_y$
by writing them as follows:
\[
\phi^{-1}\phi_x \approx \frac{1}{h} \phi_{i+1,j}^T(\phi_{i+1,j} -
\phi_{i,j}) \qqand \phi^{-1}\phi_y \approx \frac{1}{h}
\phi_{i,j+1}^T(\phi_{i,j+1} - \phi_{i,j}),
\]
where $\phi_{i,j}^T$ is the transpose of the matrix $\phi_{i, j}$.
Substituting this into (\ref{HMlag}) yields the following discrete
Lagrangian (up to an unimportant constant):
\[
  \calL_d = - \frac{1}{h^2} \tr(\phi_{i,j}^T\phi_{i+1,j}) -
  \frac{1}{h^2} \tr(\phi_{i,j}^T\phi_{i,j+1}).
\]

In order to ensure that $\phi_{i,j} \in SO(n)$, we need to impose the
constraint that $\phi_{i,j}^T \phi_{i,j} = I$.  We are thus led to
consider the following action:
\begin{equation} \label{HMaction}
  S(\phi) = \sum_{i,j} \left( \tr(\phi_{i,j}^T\phi_{i+1,j}) +
    \tr(\phi_{i,j}^T\phi_{i,j+1}) - \frac{1}{2}
    \tr\left(\Lambda_{i,j}(\phi_{i,j}^T\phi_{i,j} - I)\right) \right),
\end{equation}
where we have rescaled the Lagrange multipliers $\Lambda_{i,j}$ to get
rid of the factor $-1/h^2$.  Note that $\Lambda_{i,j}$ is a symmetric
matrix.

The field equations are obtained by demanding that $S$ be stationary
under arbitrary variations; they are given by 
\begin{equation} \label{FE}
  \phi_{i+1,j} + \phi_{i-1,j} + \phi_{i,j+1} + \phi_{i,j-1} =
  \phi_{i,j}\Lambda_{i,j}.
\end{equation}
We multiply these equations by $\phi_{i,j}^T$ from the right, and use
the fact that $\phi_{i,j} \Lambda_{i,j} \phi^T_{i,j}$ is a symmetric
matrix in order to get rid of the Lagrange multipliers:
\begin{multline*}
  \phi_{i+1,j}\phi_{i,j}^T + \phi_{i,j+1}\phi_{i,j}^T +
  \phi_{i-1,j}\phi_{i,j}^T + \phi_{i,j-1}\phi_{i,j}^T = \\
  \phi_{i,j}\phi_{i+1,j}^T + \phi_{i,j}\phi_{i,j+1}^T +
  \phi_{i,j}\phi_{i-1,j}^T + \phi_{i,j}\phi_{i,j-1}^T.
\end{multline*}

By introducing the following quantities, 
\[
m_{i+1,j} = \phi_{i+1,j}\phi_{i,j}^T - \phi_{i,j}\phi_{i+1,j}^T \qqand
n_{i,j+1} = \phi_{i,j+1}\phi_{i,j}^T - \phi_{i,j}\phi_{i,j+1}^T,
\]
the field equations can be rewritten as the following set of
conservation laws:
\begin{equation}
  m_{i+1,j} + n_{i,j+1} = m_{i,j} + n_{i,j}.  \label{conslaw}
\end{equation}

Finally, let us introduce the \emph{discrete momenta} $M_{i,j}$ and
$N_{i,j}$, defined as
\[
  M_{i,j} = \phi_{i-1,j}^T m_{i,j} \phi_{i-1,j} \qqand N_{i,j} =
  \phi_{i,j-1}^T n_{i,j} \phi_{i,j-1}.
\]
The field equations governing the behaviour of these quantities are
then easily determined to be, on the one hand 
\begin{equation} \label{MV1}
\begin{cases}
  M_{i,j} = \alpha_{i,j} - \alpha_{i,j}^T & \text{where $\alpha_{i,j}
    = u_{i-1,j}$}; \\
  N_{i,j} = \beta_{i,j} - \beta_{i,j}^T & \text{where $\beta_{i,j} =
    v_{i,j-1}$},
\end{cases}
\end{equation}
as well as, on the other hand, the counterpart of (\ref{conslaw}):
\begin{equation}
  \label{MV2}
  M_{i+1,j} + N_{i,j+1} = \Ad_{\alpha_{i,j}^T} M_{i,j} +
  \Ad_{\beta_{i,j}^T} N_{i,j}.
\end{equation}
The similarities with the Moser-Veselov equations for the discrete
rigid body are obvious (compare with equation~($4$) in
\cite{MoserVeselov91}). 

\begin{remark}  \label{remark:equivalence}
It is now straightforward to see the equivalence between the
Moser-Veselov and the Euler-Poincar\'e equations.  Indeed, starting
from the reduced Lagrangian $l$, put
\[
  M_{i+1, j} = R^\ast_{u_{i, j}} \d l(\cdot, v_{i,j}) \qqand N_{i,
    j+1} = R^\ast_{v_{i, j}} \d l(u_{i,j}, \cdot),
\]
which can be interpreted as a \emph{discrete Legendre transformation}
\cite{discreet}.  Furthermore, put $\alpha_{i,j} = u_{i-1, j}$ and
$\beta_{i,j} = v_{i,j-1}$.  The Euler-Poincar\'e equations
(\ref{eq:EP}) then reduce to the Moser-Veselov equations derived
above.
\end{remark}

\subsection{Relation with the Euler-Poincar\'e equations}

Instead of deriving the Euler-Poincar\'e equations from the
Moser-Veselov equations, we can also start directly from the discrete
action sum (\ref{HMaction}) and proceed as in section~\ref{sec:DEP}.

The unreduced field equations are given by (\ref{FE}).  We now
multiply these equations from the \emph{left} (rather than from the
right as in the derivation of the Moser-Veselov equations) to obtain
the following set of discrete Euler-Poincar\'e equations:
\begin{equation} \label{EPharm}
  u_{i,j} + v_{i,j} + u_{i-1,j}^T + v_{i,j-1}^T = \Lambda_{i,j},
\end{equation}
together with the integrability condition 
\begin{equation} \label{integrab}
  u_{i,j}v_{i+1,j}u_{i,j+1}^{-1}v_{i,j}^{-1} = e.
\end{equation}
Using then the symmetry of $\Lambda_{i,j}$, we eliminate the
multipliers $\Lambda_{i,j}$ to arrive at the following expression:
\[
u_{i,j} + v_{i,j} - u_{i-1,j} - v_{i,j-1} = u_{i,j}^T + v_{i,j}^T -
u_{i-1,j}^T - v_{i,j-1}^T.
\] 
If we view $\varphi$ as a $\gl(n)$-valued discrete one-form in the
sense of \S\ref{sec:discDG}$\,\ref{sec:discform}\,$, then the
Euler-Poincar\'e equations can be conveniently expressed using the
discrete codifferential:
\begin{equation} \label{eqcodiff}
  \delta \varphi = (\delta\varphi)^T, 
\end{equation}
or $\AS{\delta \varphi} = 0$, where $\AS{\cdot}$ denotes the
antisymmetric part of a matrix: $\AS{A} = \frac{1}{2}(A - A^T)$.  Many
authors (see, for instance, \cite{wood2, MarcoLMP}) express the
continuum equations for harmonic mappings taking values in a Lie group
$G$ with bi-invariant metric as $\d \alpha = 0 = \delta \alpha$, where
$\alpha$ is a $\gothg$-valued one-form on the base space.  Equations
(\ref{integrab}) and (\ref{eqcodiff}) are the discrete counterpart of
these continuum equations.

\section{Generalizations} \label{sec:general}

\subsection{Field theories on non-trivial base
  spaces} \label{sec:nontrivial} 

Up until now, we considered discretizations of field theories with two
independent variables, i.e. where the base space is $\bfR^2$.  From
the beginning of the paper, we discretized $\bfR^2$ by considering the
square mesh $(V, E, F)$ in $\bfR^2$ whose vertices have integer
coordinates.  Other discretisations of $\bfR^2$ can be introduced as
in \cite{discreet} by noting that $(V, E, F)$ is a planar graph; the
theory of this paper can be readily generalized to the case of
arbitrary planar graphs.

Secondly, the generalization to an arbitrary base space $X$, not
necessarily $\bfR^2$, is straightforward under the assumption that a
simplicial complex can be embedded in $X$.  Let us now briefly discuss
this case.

As before, unreduced and reduced fields are maps associating a group
element to each $0$-dimensional and $1$-dimensional simplex,
respectively.  Other generalizations are also possible, for instance
where one considers fields that associate group elements to
higher-dimensional simplices, but these types of fields do not arise
in Euler-Poincar\'e reduction and cannot be treated with the Lie
groupoid framework of section~\ref{sec:groupoids}.  For an
introduction to such field theories, see \cite{Wise}.

The theory of discrete differential forms starting from a simplicial
complex and its dual is developed to great detail in \cite{DEC,
  HiraniPhd} for spaces of arbitrary dimension and topology.  To study
discrete Lagrangian field theories, one now has to focus on the
Cartesian products $G^{\times (n + 1)}$ and $G^{\times n}$, where $n +
1$ is the dimension of the base space.  The usual procedures of
discrete Lagrangian field theories and discrete reduction of the
previous sections carry through to this case without significant
changes.

A significant change occurs when the base space is not simply
connected.  In that case, proposition~\ref{prop:holonomy} is no longer
valid.  Consequently, a reduced field then no longer induces a
uniquely determined unreduced field.  This is similar to the
continuous case; see \cite{MarcoLMP}.

\subsection{Discrete field theories on Lie
  groupoids} \label{sec:groupoids}

For the better part of this paper, we have considered discrete field
theories associating group elements to either the vertices (unreduced
fields), or to the edges of the mesh (reduced fields).  It turns out
that both are special cases of discrete field theories taking values
in \emph{Lie groupoids}.  This point of view was introduced in
\cite{discreet}, to which we refer for a detailed overview.  In this
section, we will briefly show how Lie groupoid field theories provide
a unified framework for Euler-Poincar\'e reduction.  We will focus
mostly on a Lie groupoid version of theorem~\ref{thm:eulerpoincare}, but
note that other constructions, such as the Noether theorem and the
Poincar\'e forms of section~\ref{sec:noether}, also have their
counterpart in the Lie groupoid framework.

For more information on Lie groupoids, as well as their role in
discrete mechanics, the reader is referred to \cite{Mackenzie,
  Weinstein96, GroupoidMech05}.

A \emph{Lie groupoid} is a set $G$ with a partial multiplication $m$,
a subset $Q$ of $G$ whose elements are called \emph{identities}, two
submersions $\alpha, \beta: G \rightarrow Q$ (called \emph{source} and
\emph{target} maps respectively), which both equal the identity on
$Q$, and an inversion mapping $i: G \rightarrow G$.  A pair of Lie
groupoid elements $(g,h)$ is said to be \emph{composable} if the
multiplication $m(g, h)$ is defined; the set of composable pairs will
be denoted by $G_2$.  We will denote the multiplication $m(g,h)$ by
$gh$ and the inversion $i(g)$ by $g^{-1}$.  In addition, these data
must satisfy the following properties, for all $g, h, k \in G$:
\begin{enumerate}
\item the pair $(g, h)$ is composable if and only if $\beta(g) =
  \alpha(h)$, and then $\alpha(gh) = \alpha(g)$ and $\beta(gh) =
  \beta(h)$;

\item if either $(gh)k$ or $g(hk)$ exists, then both do, and they are
  equal;

\item $\alpha(g)$ and $\beta(g)$ satisfy $\alpha(g)g = g$ and
  $g\beta(g) = g$;

\item the inversion satisfies $g^{-1}g = \beta(g)$ and $gg^{-1} =
  \alpha(g)$.
\end{enumerate}

A Lie group $G$ can be considered as a Lie groupoid over a singleton
$\{e\}$: the source and target maps $\alpha$ and $\beta$ map any group
element onto $e$ and the multiplication is defined everywhere.

Another example is the pair groupoid $Q \times Q$, where $Q$ is a
manifold.  The pair groupoid is a Lie groupoid over $Q$, and the
source and target mappings are defined as follows: $\alpha(q_0, q_1) =
q_0$, and $\beta(q_0, q_1) = q_1$.  The multiplication of composable
elements is then given by $(q_0, q_1) (q_1, q_2) = (q_0, q_2)$.  The
set of units in $Q \times Q$ is $\{(q, q): q \in Q\}$.

A \emph{groupoid morphism} is a pair of maps $\phi: G \rightarrow
G'$ and $f: Q \rightarrow Q'$ satisfying $\alpha' \circ \phi = f \circ
\alpha$, $\beta' \circ \phi = f \circ \beta$ and such that $\phi(gh) =
\phi(g)\phi(h)$ whenever $(g, h)$ is composable.  Note that $(\phi(g),
\phi(h))$ is a composable pair whenever $(g,h)$ is composable.

\subsubsection{Discrete fields}

The set of edges $E$ is a subset of the pair groupoid $V \times V$,
but it is not a groupoid in itself because the multiplication of two
elements of $E$ is not necessarily again an element of $E$.
Nevertheless, it turns out that much can be gained from considering
$E$ as a ``local groupoid'' in the sense of \cite{discreet}, rather
than just as a subset of $V \times V$.  By a ``local groupoid'', we
mean here that, even if the multiplication is not defined in $E$, one
can still define \emph{composable edges} as pairs of edges $\gothe_1,
\gothe_2$, such that $\beta(\gothe_2) = \alpha(\gothe_1)$, where
$\alpha(x_0, x_1) = x_0$ and $\beta(x_0, x_1) = x_1$, i.e. $\alpha$
and $\beta$ are the usual source and target mappings of the pair
groupoid $V \times V$, restricted to $E$.  Hence, $E$ is a groupoid in
all aspects but one, the multiplication.

With the conventions introduced above, we have for $\gothe = (x_0,
x_1)$ that $\gothe^{-1} = (x_1, x_0)$.  In addition to the mesh $(V,
E)$ in $\bfR^2$, we now consider an arbitrary Lie groupoid $\Gamma$
over a manifold $Q$.  The idea is to define \emph{discrete fields} as
mappings from the ``local groupoid'' $E$ to the Lie groupoid $\Gamma$.
In particular, if $\varphi$ is a discrete field, then $\varphi$ maps
composable edges in $E$ to composable elements in $\Gamma$, and,
secondly, if $\gothe$ is any edge in $E$, then $\varphi(\gothe^{-1}) =
\varphi(\gothe)^{-1}$.  This is worked out in more detail in the
following definition:

\begin{definition} \label{def:discfield}
  A \emph{discrete field} is a pair $\varphi = (\varphi_{(0)},
  \varphi_{(1)})$, where $\varphi_{(0)}$ is a map from $V$ to $Q$ and
  $\varphi_{(1)}$ is a map from $E$ to $\Gamma$ such that
  \begin{enumerate}
    \item $\alpha(\varphi_{(1)}(x, y)) = \varphi_{(0)}(x)$ and
      $\beta(\varphi_{(1)}(x, y)) = \varphi_{(0)}(y)$;
    \item for each $(x, y) \in E$, $\varphi_{(1)}(y, x) =
      [\varphi_{(1)}(x, y)]^{-1}$;
    \item for each $x \in V$, $\varphi_{(1)}(x, x) = \varphi_{(0)}(x)
      \in Q$.
  \end{enumerate}
\end{definition}

Even though we are working with an object $E$ which is not quite a
groupoid, it turns out that any discrete field $\varphi: E \rightarrow
\Gamma$ may be extended to an actual groupoid morphism from $V \times
V$ to $\Gamma$.  This can be shown quite easily; for a proof, we refer
to proposition~7 in \cite{discreet}.  Note that
proposition~\ref{prop:holonomy} is a special case of this extension
property. 

\begin{example}
  Consider a discrete field $(\varphi_{(0)}, \varphi_{(1)})$ taking
  values in the pair groupoid $Q \times Q$.  Because of the conditions
  in definition~\ref{def:discfield}, the map $\varphi_{(1)}: V \times
  V \rightarrow Q \times Q$, is equal to $\varphi_{(0)} \times
  \varphi_{(0)}$.  In this case, the discrete field is completely
  specified once we are given $\varphi_{(0)}: V \rightarrow Q$.  For
  $Q = G$, this corresponds to the class of unreduced fields studied
  in section~\ref{sec:DEP}.
\end{example}

\begin{example}
  Consider now the case where the Lie groupoid $\Gamma$ is a Lie group
  $G$ and let $(\varphi_{(0)}, \varphi_{(1)})$ be a discrete field.
  As $\varphi_{(0)}(x) = e$ for all $x \in V$, a discrete field can be
  identified with a map $\varphi_{(1)}: V \times V \rightarrow G$.  As
  in section~\ref{sec:reconstruction}, this is a flat discrete
  connection, or, equivalently, a reduced field.
\end{example}

\subsubsection{Reduction}

The previous two examples show that the Lie groupoid framework
encompasses both the case of unreduced and reduced fields (when
$\Gamma = G \times G$ or $\Gamma = G$, respectively).  The following
theorem shows that the Euler-Poincar\'e reduction of one Lie groupoid
field theory yields another one:

\begin{theorem}[see \cite{discreet}] \label{thm:red}
  Let $\Gamma'$ be a Lie groupoid over a manifold $Q'$ and consider a
  morphism $(\Phi, f): (\Gamma, Q) \rightarrow (\Gamma', Q')$.
  Furthermore, let $L': \bfG'^k \rightarrow \bfR$ be a Lagrangian on
  $\bfG'^k$ and consider the induced Lagrangian $L = L' \circ \Psi$ on
  $\bfG^k$, where $\Psi: \bfG^k \rightarrow \bfG'^k$ is the map
  associated to $\Phi$.

  A morphism $\phi : V \times V \rightarrow \Gamma$ will satisfy the
  discrete field equations for $L$ if the induced morphism $\Phi \circ
  \phi: V \times V \rightarrow \Gamma'$ satisfies the discrete field
  equations for $L'$.
\end{theorem}

\begin{lemma} \label{lemma:submersion}
  If the morphism $(\Phi, f)$ in theorem~\ref{thm:red} is a
  submersion, then a discrete field $\phi: V \rightarrow V \rightarrow
  \Gamma$ will satisfy the discrete field equations for $L$ if and
  only if the induced field $\Phi \circ \phi: V \times V \rightarrow
  \Gamma'$ satisfies the field equations for $L'$.
\end{lemma}
\begin{mproof}
  The proof follows that of Corollary~4.7 in \cite{GroupoidMech05}.
\end{mproof}

A few remarks are in order here.  The manifold $\bfG^k$ consists of
sequences of $k$ composable elements $(g_1, g_2, \cdots, g_k)$ in
$\Gamma$, such that $g_1 \cdot g_2 \cdots g_k$ is a unit:
\begin{align*}
  \bfG^k = \{ (g_1, g_2, \ldots, g_k) \in G^{\times k} : &\,\, \alpha(g_{i +
    1}) = \beta(g_i), \alpha(g_1) = \beta(g_k) \qqand \\
   & g_1 \cdot g_2 \cdots g_k = e_{\alpha(g_1)} \}.
\end{align*}
The manifold $\bfG^k$ is the manifold where the Lagrangian is defined,
and as such it is the generalization to the case of Lie groupoids of
the Cartesian products $G \times G \times G$ and $G \times G$ used in
section~\ref{sec:disclagrangian}.  It is straightforward to check
that, for $k = 3$, $\bfG^3 = G \times G \times G$ when $\Gamma = G
\times G$, and $\bfG^3 = G \times G$ for $\Gamma = G$.

The map $\Phi: G \times G \rightarrow G$, introduced in
section~\ref{sec:disclagrangian}, and defined as $\Phi(g,g') =
g^{-1}g'$, is a Lie groupoid morphism.  When we put $\Gamma = G \times
G$ and $\Gamma' = G$, and use $\Phi$ in theorem~\ref{thm:red}, we
obtain the reduction theorem~\ref{thm:eulerpoincare}.  In contrast to
discrete mechanics, it is not very likely that a reconstruction
theorem can be proved in the context of Lie groupoids; in the case of
Lie groups, for instance, there is the obstruction of flatness.

\section{Concluding remarks}

In this paper, we have shown that the well-known concepts of symmetry
and Euler-Poincar\'e reduction can be extended by very modest means to
the case of discrete field theories.  It turned out that many of the
constructions known from continuum field theories have a natural
discrete counterpart.  However, much remains to be done.

Throughout the text, we remarked that almost all of our constructions
can be extended to the case of field theories taking values in an
arbitrary Lie groupoid $\Gamma$.  Such a framework might be useful in
the case of general Lagrangian reduction.  In addition, as we pointed
out before, many of the definitions introduced here gain much in
clarity when rephrased in this more general language.

As a second point of interest, we are also interested in the concept
of multisymplecticity.  Because of their variational nature, it is to
be expected that the Euler-Poincar\'e equations are multisymplectic in
the sense of \cite{MPS98, Bridges01}.  However, as the focus of this
article was on theoretical developments rather than on the
construction of practical integration schemes, we have chosen to leave
this topic for future work.

\section*{Acknowledgements}

The author is a Research Assistant of the Research Foundation ---
Flanders (Belgium), whose financial support is gratefully
acknowledged.  I would like to thank Frans Cantrijn for useful
discussions and a critical reading of this manuscript.  Furthermore,
I am grateful to an anonymous referee for valuable criticisms which
significantly improved both the contents and the presentation of this
paper. 

% \bibliographystyle{ams-abbrv-jvk}
% \bibliography{biblio}

\end{document}